\documentclass[fleqn,usenatbib]{mnras}

\usepackage{newtxtext,newtxmath}
\usepackage[T1]{fontenc}

\usepackage{multirow}

\DeclareRobustCommand{\VAN}[3]{#2}
\let\VANthebibliography\thebibliography
\def\thebibliography{\DeclareRobustCommand{\VAN}[3]{##3}\VANthebibliography}

\usepackage{graphicx}
\usepackage{url}
\usepackage{amsmath,siunitx}

\title[4XJ1751-2759 - Magnetar candidate?]{4XMM J175136.8-275858: A New Magnetar Candidate?}

\author[R. Webbe et al.]{
Robbie Webbe,$^{1}$\thanks{E-mail: robbie.webbe@irap.omp.eu} \thanks{\protect\url{https://orcid.org/0000-0003-1689-3723}}
Norman Khan,$^{1}$\thanks{\protect\url{https://orcid.org/0000-0002-3955-0697}}
N. A. Webb$^{1}$
and E. Quintin$^{1,2}$\thanks{\protect\url{https://orcid.org/0000-0002-7116-2897}}
\\
$^{1}$Institut de Recherche en Astrophysique et Planétologie (IRAP), Avenue du Colonel Roche, 31028 Toulouse\\
$^{2}$European Space Agency (ESA), European Space Astronomy Centre (ESAC), Camino Bajo del Castillo s/n, 28692 Villanueva de la Cañada, Madrid, Spain\\
}

\date{Accepted XXX. Received YYY; in original form ZZZ}

\pubyear{\the\year{}}

\begin{document}
\label{firstpage}
\pagerange{\pageref{firstpage}--\pageref{lastpage}}
\maketitle

\begin{abstract}
Magnetars are very rare astrophysical objects, with $\sim$31 known to date. They are best understood as highly magnetised neutron stars, but a greater number need to be found to constrain their role in stellar evolution pathways. We apply a novel approach for the detection of fast, transient X-ray sources, using a revised version of the EPIC XMM-Newton Outburst Detector (EXOD) with the aim of detecting and identifying new and rare variable compact objects. We detect a transient, variable source notable for its strong variability and hard spectrum. The emission from 4XMM J175136.8-275858 is well characterised by a blackbody, with temperatures between $\sim$1.8--5\,keV during its lower luminosity phase. Its temperature is poorly constrained during its brightest phase, and we observe an increase in luminosity by two orders of magnitude over timescales of a few ks. This is driven by increased emission of X-rays at energies above 2\,keV, with a luminosity decay potentially over weeks or months. Derived luminosities for 4XJ1751-2759 range up to $\sim10^{35} \text{\,erg s}^{-1}$ at 8\,kpc at the Galactic centre, but neutral hydrogen column densities are greater than predicted Galactic values possibly implying a greater distance to the source, still within our galaxy, further increasing its luminosity. A consideration of optical and IR information in combination with the X-ray observations allow us to exclude the possibility that 4XJ1751-2759 is a star, rotationally powered pulsar or supergiant fast X-ray transient. This rapid, hard, variability is closest to that of outbursts in magnetars than any other known class of X-ray transient.

\end{abstract}


%%%%%%%%%%%%%%%%%%%%%%%%%%%%%%%%%%%%%%%%%%%%%%%%%%

%%%%%%%%%%%%%%%%% BODY OF PAPER %%%%%%%%%%%%%%%%%%

\section{Introduction}
\label{sec:intro}

Magnetars are a rare sub-class of neutron stars which are believed to be highly magnetised and young neutron stars. Their energetic emission is powered by the decay of their strong magnetic fields as opposed to being primarily rotationally powered, and is expected to last 1--10 kyr \citep{kaspi_magnetars_2017}. The braking (spin-down) caused by these strong magnetic fields, typically with $B > 5\times10^{13}$\,G \citep{kaspi_magnetars_2017}, also leads to long pulsation periods, of the order of $\sim2-10$\,s, with larger period derivatives than for typical pulsars, see Figure 10 of \citet{olausen_mcgill_2014} for a comparison of timing measures of magnetars and rotationally powered pulsars. The current population has been discovered primarily due to the detection of high energy phenomena, Soft Gamma Ray Repeaters (SGRs) and Anomalous X-ray Pulsars (AXPs). Both SGRs and AXPs show energetic emission at high energies, gamma- and X-rays respectively. SGRs show repeated bursting behaviour \citep[][etc.]{norris_nature_1991}, and AXPs show energetic emission at X-ray energies \citep[][etc.]{mereghetti_anomalous_1999}, and in both cases they cannot be explained purely by the spin-down of rotationally powered pulsars. At present there are a very small number of such objects, with only 24 confirmed and 6 candidate sources\footnote{\url{https://www.physics.mcgill.ca/~pulsar/magnetar/main.html}}. The majority of known magnetars (17 of 31) are located at heights close to the Galactic plane, at elevations of less than 100pc \citep{olausen_mcgill_2014}. When not showing significant variability, magnetars are classified by their  quiescent X-ray luminosity as either 'persistent' ($L_X \geq 10^{33}$ erg s$^{-1}$) or 'transient' ($L_X \leq 10^{33}$ erg s$^{-1}$) \citep{kaspi_magnetars_2017}. For a full review of the population of magnetars and their characteristics, see \cite{kaspi_magnetars_2017}, \cite{olausen_mcgill_2014}, and the references therein.

Magnetars display both long and short term, high amplitude, variability in X-ray energy ranges. There are three classes of X-ray variability phenomena observed to date in magnetars: bursts; giant flares; outbursts. This variability is observed in addition to familiar pulsar variability phenomena like glitches. Bursts, which have the shortest duration, occur on millisecond to second timescales, and have been observed in several magnetars \citep[e.g.][]{gogus_statistical_2000,sakamoto_probing_2011,van_der_horst_sgr_2012,gogus_magnetar-like_2016}. These bursts are often clustered together in time, interspersed with extended periods of quiescence, and the bursts have peak X-ray luminosities in the range from $\sim10^{36} - 10^{43}\,\text{\,erg s}^{-1} $. On longer timescales, the giant flares in SGR 0526-66 \citep{evans_location_1980,cline_detection_1980}, SGR 1806-20 \citep{hurley_exceptionally_2005}, and SGR 1900+14 \citep{hurley_giant_1999} have been observed to last hundreds of seconds. Their peak luminosities of up to $\sim10^{47} \text{\,erg s}^{-1}$ are far greater than those observed during the shorter bursts and to date only three such giant flares have been observed. These luminosities imply that bursts and giant flares can reach well above the Eddington limit for neutron stars ($\sim 10^{38}\ \text{erg s}^{-1}$). The origin of the bursts and flares is still uncertain, but possible mechanisms for triggering bursts and flares include quakes in the crust of the neutron star, magnetohydrodynamic instabilities in the stellar core, and magnetic reconnection in the magnetosphere \citep[e.g.][]{thompson_soft_1995,thompson_giant_2001,lyutikov_explosive_2003}.

Outbursts have been observed in some magnetars \citep[e.g.][]{kaspi_major_2003,tam_x-ray_2008,israel_2008_2010}, and are characterised by initial large increases in luminosity on timescales of seconds followed by bright phases that last significantly longer than those seen in giant flares. During the initial rise in luminosity, a correlation between hardness ratio and flux has been observed in several sources \citep[e.g.][]{scholz_post-outburst_2012,dib_rxte_2012,an_spectral_2013}, with typical values for the 4-10\,keV to 2-4\,keV ratio increasing from below 0.4 to over 3 as the sources reach their brightest. The prolonged bright phases can last for months or years, where the X-ray luminosity persists above the previously observed quiescent levels at factors of greater than 10-1000. A review of outbursts conducted by \cite{coti_zelati_systematic_2018} suggests that there may be an upper limit to the luminosity during these outbursts of $\sim10^{36} \text{\,erg s}^{-1}$, with an anti-correlation between outburst amplitude and the quiescent luminosity.

The discovery of more magnetars is key to our understanding of these sources. With a larger population of magnetars we could constrain the pathways which lead to their production, further constrain the neutron star equations of state, and understand what proportion of neutron stars exist with these extreme magnetic fields \citep{kaspi_radio_2016}. A larger population would also allow us to constrain the evolution of the magnetic fields, and thus the potential range of lifetimes of such sources \citep[e.g.][]{mondal_life_2021}. This would further enable us to understand the production rates, not only of magnetars but of all neutron stars. This is in addition to the ability to monitor such sources for further bursts, outbursts and giant flares for potential fast follow-up and analysis, and monitoring these transient outbursts will help to further constrain the neutron star equation of state .

The variable and transient nature of magnetars make them ideal candidates for detection with transient searches in X-ray data in both new and archival data. The EPIC XMM-Newton Outburst Detector \citep[EXOD,][Khan et al. in prep]{pastor-marazuela_exod_2020} has been developed to search for this type of rapid transient behaviour from even the faintest X-ray sources. This is particularly true for transient magnetars which may be below detection thresholds while in quiescence but detectable when in outburst. Magnetar outbursts are generally fairly hard and using a revised version of EXOD that allows to search for bursts during  what is traditionally considered 'bad time intervals' (Khan et al. in prep.), increases the potential of the archival data. During preliminary testing of EXOD several variable sources were detected, and this article will focus on one particular source of interest, a magnetar candidate.

\section{Data \& Methods}
\label{sec:methods_data}

\subsection{Observational data}
\label{subsec:methods_data-obs}
The X-ray source 4XMM J175136.8-275858, henceforth 4XJ1751-2759, was detected in a pair of \emph{XMM-Newton} observations of the Galactic plane taken as part of a multi-year heritage programme on the 8th October 2022. It was detected as a 4XMM-DR13 source with the EPIC MOS1 and MOS2 cameras during the first observation, 0886120901, but was outside the field of view of the pn detector. The source was then detected with all three cameras during the second observation, 0886121001. 4XJ1751-2759 is in the \emph{XMM} Serendipitous Source Catalogue \citep{webb_xmm-newton_2020}, version 13, as 4XMM J175136.8-275858. 4XJ1751-2759 was then detected in a third observation taken nearly six months later on 4th April 2023, and was visible in all three cameras once again, albeit falling on a chip gap on the pn detector. The details of these observations are listed in Table \ref{tab:all_obs}. 

The outburst was identified as a fast transient (5$\sigma$) using the EPIC XMM-Newton Outburst Detector \citep[EXOD,][Khan et al. in prep]{pastor-marazuela_exod_2020} with a temporal binning of $50 \ \mathrm{s}$ in the $0.5 - 12.0 \ \mathrm{keV}$ energy band. The source was then flagged as being of interest, and scheduled for expert examination, during post-processing of the lightcurve produced by EXOD using the approach developed by \cite{webbe_searching_2023} for the detection of Quasi-Periodic Eruptions in active galactic nuclei.

Following the identification of repeated flaring behaviour seen in the \emph{XMM-Newton} pointed observations we then conducted a broader search for observations of the location in the archives of high energy observatories (see Fig. \ref{fig:flux_evol2}). There were ten \emph{XMM-Newton} slew observations which should include 4XJ1751-2759 within the field of view, but it was only detected as a source during one of these, on 30th March 2017, with upper limits available on the other nine obtained through the HILIGT\footnote{\url{http://xmmuls.esac.esa.int/hiligt/}} upper limit server. Three further observations which included the source position were found in archival \emph{Chandra} observations from May to July 2022. Across these observations the source is detected by \texttt{wavdetect}, as part of the \texttt{ciao}\footnote{\url{https://cxc.cfa.harvard.edu/ciao/}} package in two of the three observations. The \emph{XMM-Newton} and \emph{Chandra} observations used in this work are listed in Table \ref{tab:all_obs}. There were several observations of the source position with the \emph{Swift} XRT, however the catalogued source, 2SXPS J175136.8-275856, is only detected in one of these observations and during this only eleven photons were detected, and as such this \emph{Swift} XRT data does not form a part of this subsequent analysis. At higher energies there are no \emph{Fermi} triggers around the time of the outburst, but we do examine the time-tagged event data from the day of the outburst as collected by the all-sky monitors. We apply the \emph{Fermi} targeted search tool\footnote{\url{https://fermi.gsfc.nasa.gov/ssc/data/analysis/gbm/}} to identify the existence (or not) of any sub-threshold triggers which could be associated with the soft X-ray outburst identified by \emph{XMM-Newton}.

\begin{table}
\caption{List of observations made with \emph{XMM-Newton} and \emph{Chandra} used in this analysis. Columns list the observatory, the observation ID, date and exposure time of observation, and whether the source was detected in the observation.}
\label{tab:all_obs}
\centering
\begin{tabular}{ccccc}
\hline\hline
Obs. & OBSID & Date & Exposure & Detect \\
 &  &  & (ks) &  \\
\hline
   Chandra & 24064 & 2022-05-26 & 4.19 & Y \\   
   Chandra & 24193 & 2022-06-28 & 4.36 & Y \\
   Chandra & 24220 & 2022-07-24 & 4.19 & N \\
   XMM & 0886120901 & 2022-10-08 & 23.0 & Y \\
   XMM & 0886121001 & 2022-10-08 & 23.0 & Y \\ 
   XMM & 0886080801 & 2023-04-03 & 23.0 & Y \\ 
\hline
\end{tabular}
\end{table}

\begin{figure}
\centering
\includegraphics[width=\columnwidth]{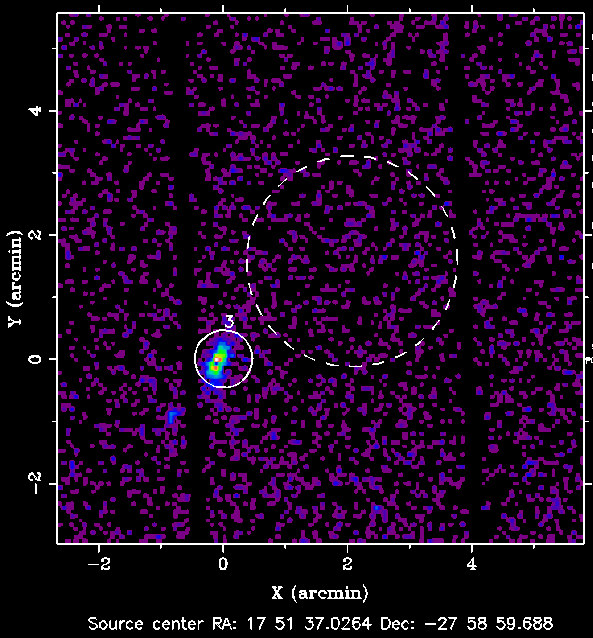}
\caption{Location of the source 4XJ1751-2759 as detected by the EPIC pn camera during observation 0886121001 with \emph{XMM-Newton}. The solid line represents the source region used by the \emph{XMM-Newton} processing pipeline, with the dashed line delineating the background region used in creating automated timing and spectral products.}
\label{fig:xmm_src_img}
\end{figure}

\begin{figure}
\centering
\includegraphics[width=\columnwidth]{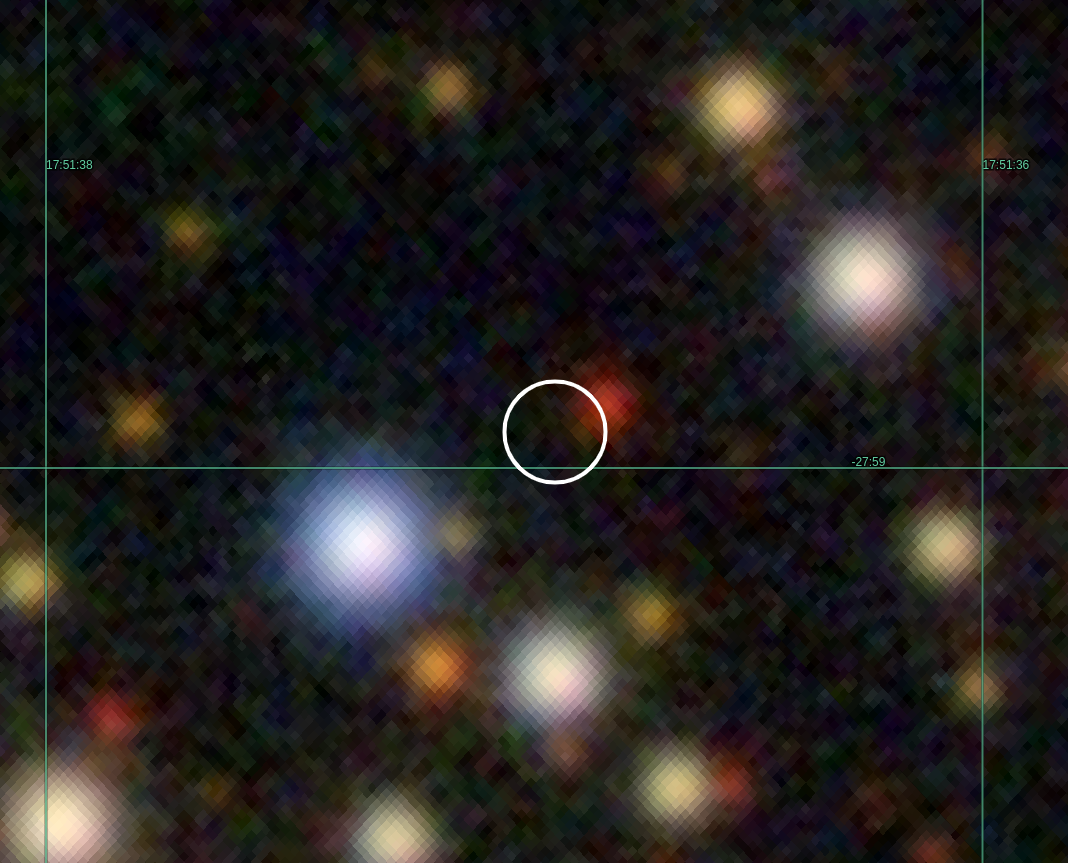}
\caption{3$\sigma$ error circle of the X-ray position of 4XJ1751-2759 (white circle), overlaid on a PanSTARRS DR1 colour image of size 30"$\times$24".}
\label{fig:opt_position}
\end{figure}

The source image, as detected during observation 0886121001 with the EPIC pn instrument and created with the \emph{XMM-Newton} pipeline processing systems\footnote{\url{https://xmm-tools.cosmos.esa.int/external/xmm_obs_info/odf/data/docs/XMM-SOC-GEN-ICD-0024.pdf}} (Rodriguez, Rosen, and Osborne 2021) is displayed in Figure \ref{fig:xmm_src_img}. The source was localised to the position 17h51m36.91s \ang{-27;58;58.98}, with a positional error of \ang{;;0.48}, at an off-axis angle in both observations of $\sim$\ang{;11;}. In Figure \ref{fig:opt_position} we present the PanSTARRS image of the position of the source, with the source location and a 3$\sigma$ positional error (\ang{;;1.43}) radius on the X-ray position overlaid, created using ESASky\footnote{\url{https://sky.esa.int/}}. There are no optical counterparts within the 3$\sigma$ X-ray position of the source.


There are no Gaia, 2MASS, or WISE catalogue sources within the 3$\sigma$ X-ray positional error. The closest optically detected source is at a distance of \ang{;;1.54}, with positional errors of less than \ang{;;0.002} in the GAIA DR3 catalogue \citep{gaia_collaboration_vizier_2022}. The closest infra-red detected sources is at a distance of \ang{;;1.54}, with a positional error of \ang{;;0.06} in the 2MASS All-Sky point source catalogue \citep{cutri_vizier_2003}. An examination of other X-ray catalogues found a possible match with the \emph{Neil Gehrels Swift Observatory} X-ray Telescope (XRT) source 2SXPS J175136.8-275856 \citep{evans_2sxps_2020} at a distance of \ang{;;2.52} with a 90\% positional error of \ang{;;3.8}. 2SXPS J175136.8-275856 was observed five times between March 2007 and April 2018, but was only detected in one of those observations on 18th March 2007.

\subsection{Temporal Analysis}
\label{subsec:methods_data-temporal}

Due to the irregular shapes, arrival times, and amplitudes of the individual flares during the outburst we are unable to adequately fit an analytical model to the shape of the multiple, overlapping, flares. As such, we instead apply a clustering of the photon event times using the \texttt{HDBSCAN} algorithm included in \texttt{sklearn}. We require a minimum of 10 photon events to constitute individual flares, and an alpha value of 0.75, to allow for the separation of individual flares. From a visual inspection of the lightcurve we select the \emph{XMM} time stamp 781623400, corresponding to approximately 8 October 2022 13:35:30 UTC, as a reference for the onset of the outburst $\sim$18.5\,ks after the observation began.

Observational data is barycentre corrected for the source location using the \texttt{barycen} tool from the \emph{XMM} Science Analysis System for \emph{XMM} observations and \texttt{axbary} for the \emph{Chandra} observation. We bin the EPIC-pn photon event data for energies in the range from 0.2--12.0\,keV at a rate of 100\,ms and bin the \emph{Chandra} event data at 5\,s due to limitations on the respective timing accuracy available, with the respective frame rates pre-binning being 73.4ms and 3.2s. We search for peak frequencies in the power spectrum for each barycentre-corrected observation using the FFT routines contained within the \texttt{stingray} module \citep{Huppenkothen2019} for \texttt{python}. We also implement the Epoch Folding and $Z^2_n$ searching algorithms from the same module for fine-level searches around peak frequencies identified in the power spectra as well as with the \texttt{HENDRICS} package \citep{bachetti_hendrics_2018}.

\section{Results}
\label{sec:results}

\subsection{Outburst of Source on 8 October 2022}
\label{subsec:outburst}

During the second observation of 4XJ1751-2759 with \emph{XMM-Newton} on 8th October 2022 (0886121001), the source was observed to flare repeatedly over a duration of approximately 1\,ks just before the end of the scheduled observation. We characterise the flaring behaviour seen during the initial rise in luminosity using the approach discussed in Section \ref{subsec:methods_data-temporal}. Through this approach we identify 15 individual flares during this outburst. We define the durations of the individual flares as being the time difference between the first and last photon event contributing to each flare (each \texttt{HDBSCAN} cluster), and these durations range from $\sim$10--100\,s. The average count rates during these flares range from $\sim$0.2--1.7\,cts s$^{-1}$, having amplitudes ranging from $\sim$20--170 times greater than that during the immediately preceding quiescent period. The second flare is the brightest of these, with a duration of $\sim$33\,s and an average count rate during this time of $\sim$1.7\,cts s$^{-1}$. There appears to be no significant structure to these flares at sub-second or multi-second timescales. As such, no conclusions can be drawn as to how closely this behaviour might reflect the bursts on sub-second timescales seen in magnetars. These flares, and their positions in the outburst, are shown in Figure \ref{fig:outburst_flares}.

\begin{figure}
\centering
\includegraphics[width=\columnwidth]{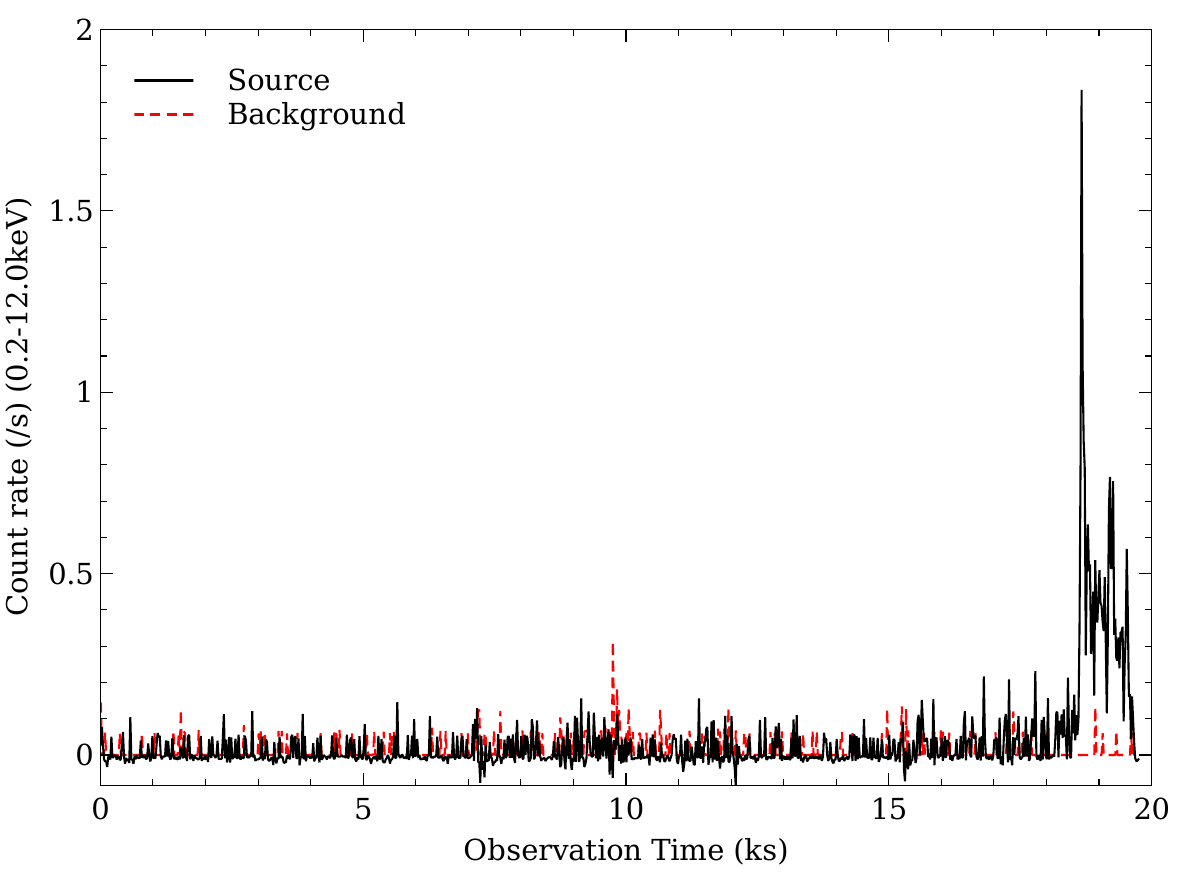}
\caption{Count rate for the energy range from 0.2--12.0\,keV as detected by the XMM-Newton EPIC pn camera during the observation 0886121001. The time is relative to that for the start of the pn camera exposure, and counts are shown in time bins of 20\,s. The black solid line shows the background-corrected source count rate, and the red dashed line shows the background count rate.}
\label{fig:full_lc}
\end{figure}

\begin{figure}
\centering
\includegraphics[width=\columnwidth]{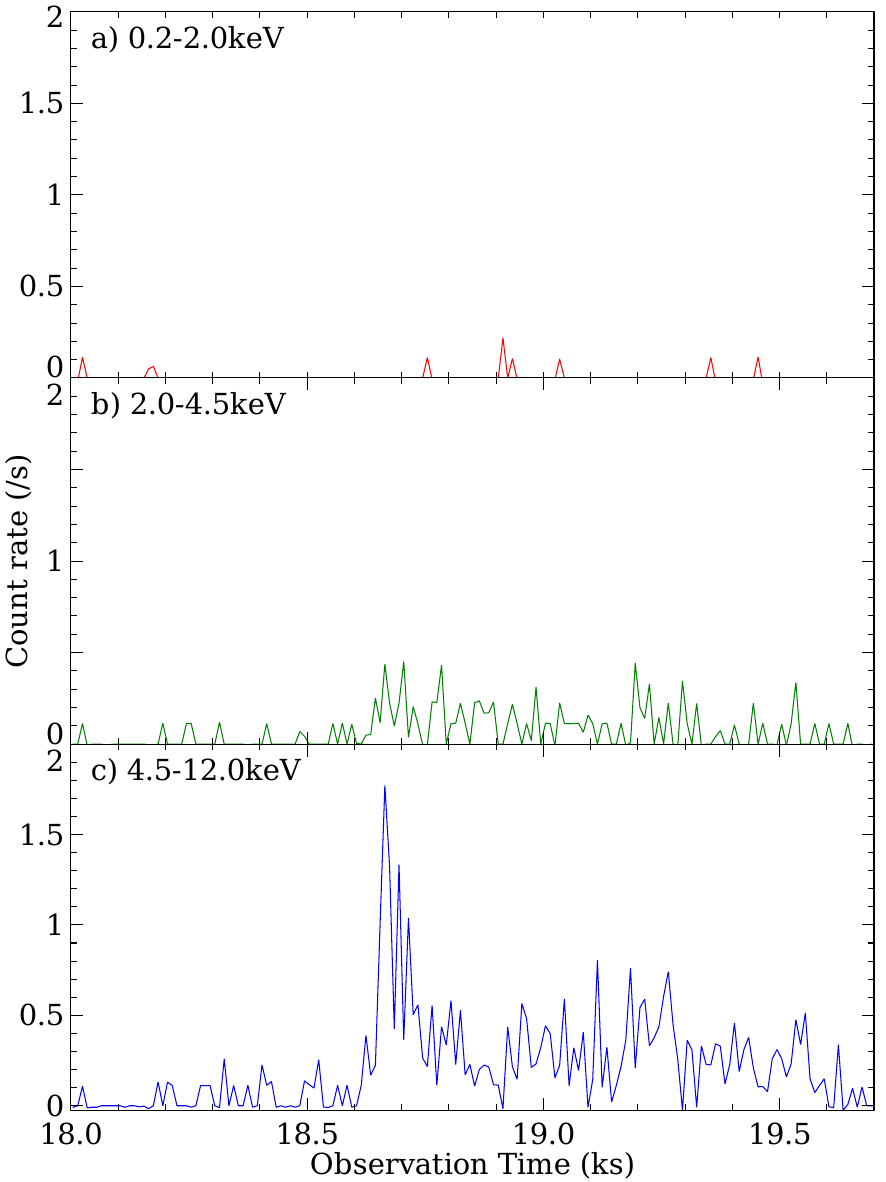}
\caption{Energy dependent behaviour during the outburst detected on 8th October 2022. Panels show the count rate in the bands (a) 0.2-2.0\,keV, (b) 2.0-4.5\,keV, and (c) 4.5-12.0\,keV. The lightcurves begin at 18\,ks following the start of the exposure time, as per the pn camera, and are binned at a rate of 10\,s.}
\label{fig:flare_lc}
\end{figure}

During the \emph{Chandra} observation 24193 there is some variability present, although it is of a smaller amplitude, and possibly longer lasting than that seen during the October 2022 outburst, with a duration of at least 3\,ks. Due to the short exposure of the observation, however, we can make no conclusions about the minimum period of any variability seen during this time, and are only constrained by the time to the next observations as to its maximum.

\subsection{Spectral evolution} 
\label{subsec:evol_spec}

Spectra are extracted for all observations, and fit against a model of an absorbed black body spectrum, \texttt{tbabs * bbodyrad}, for comparison with \cite{olausen_mcgill_2014}. Fits are performed using the \texttt{xspec} \citep{arnaud_xspec_1996} spectral fitting package, for X-ray energies in the range from 0.5--10.0\,keV, to match the wider energy range used by \citet{olausen_mcgill_2014} used to characterise magnetars. For each of the three \emph{XMM-Newton} observations we combine the spectra from all instruments which have the source in its field of view for that individual observation. Spectra are binned to ensure that there is one count in each bin and parameters are determined by minimising the Cash statistic \citep{cash_parameter_1979}. There were insufficient counts in the \emph{Chandra} observations 24064 and 24220 to fit reliably. The results of the fits to the other four observations, including the estimated unabsorbed flux in the energy range from 2--10\,keV calculated by \texttt{xspec}, are reported in Table \ref{tab:spec_fits}. For the flux estimations we apply the hydrogen column densities as determined by the fitting with \texttt{xspec}.

\begin{table*}
\renewcommand{\arraystretch}{1.5}
\caption{Spectral fits to observations of 4XJ1751-2759 with the \texttt{xspec} model \texttt{tbabs * bbodyrad}, and associated 90\% confidence intervals. The unabsorbed fluxes reported in the final column are dependent on the column densities and temperatures determined during the model fitting, and are as derived using \texttt{cflux}. The final line reports the column density determined by simultaneous fitting to all observations.}
\label{tab:spec_fits}
\centering
\begin{tabular}{cccccccc}
\hline\hline
Observation & Days after outburst & nH & $kT$ & $R_{bb}$ & $N$ & $C$ & Unabs. Flux (2--10\,keV) \\
 & & ($10^{22}$ cm$^{-2}$) & (keV) & (km) & & & ($10^{-12} \text{\,erg s}^{-1} \text{cm}^{-2}$)\\
\hline  
   24193 & -102.5 & $5.71^{+2.14}_{-1.73}$ & $1.80^{+1.44}_{-1.03}$ & $0.21^{+0.10}_{-0.07}$ & 276 & 215.2 & 7.59 \\
   0886120901 & -0.5 & $0.89^{+28.84}_{-\infty}$ & $3.99^{+\infty}_{-\infty}$ & $0.01^{+0.24}_{-0.01}$ & 211 & 187.4 & 0.316 \\ 
   0886121001 (Quiescent) & 0.0 & $0.59^{+1.29}_{-\infty}$ & $4.60^{+8.51}_{-1.82}$ & $0.02^{+0.01}_{-0.01}$ & 435 & 333.6 & 0.550 \\ 
   0886121001 (Outburst) & 0.0 & $5.08^{+2.00}_{-1.62}$ & $10.51^{+\infty}_{-\infty}$ & $0.06^{+0.04}_{-\infty}$ & 529 & 383.6 & 24.1 \\ 
   0886080801 & 177.2 & $3.03^{+1.12}_{-0.91}$ & $1.91^{+0.30}_{-0.24}$ & $0.06^{+0.01}_{-0.01}$ & 750 & 550.8 & 0.676 \\
   All & -- & $3.86^{+0.88}_{-0.75}$ & -- & 2201 & 1685.9 & -- \\
\hline
\end{tabular}
\end{table*}    

\begin{figure}
\centering
\includegraphics[width=\columnwidth]{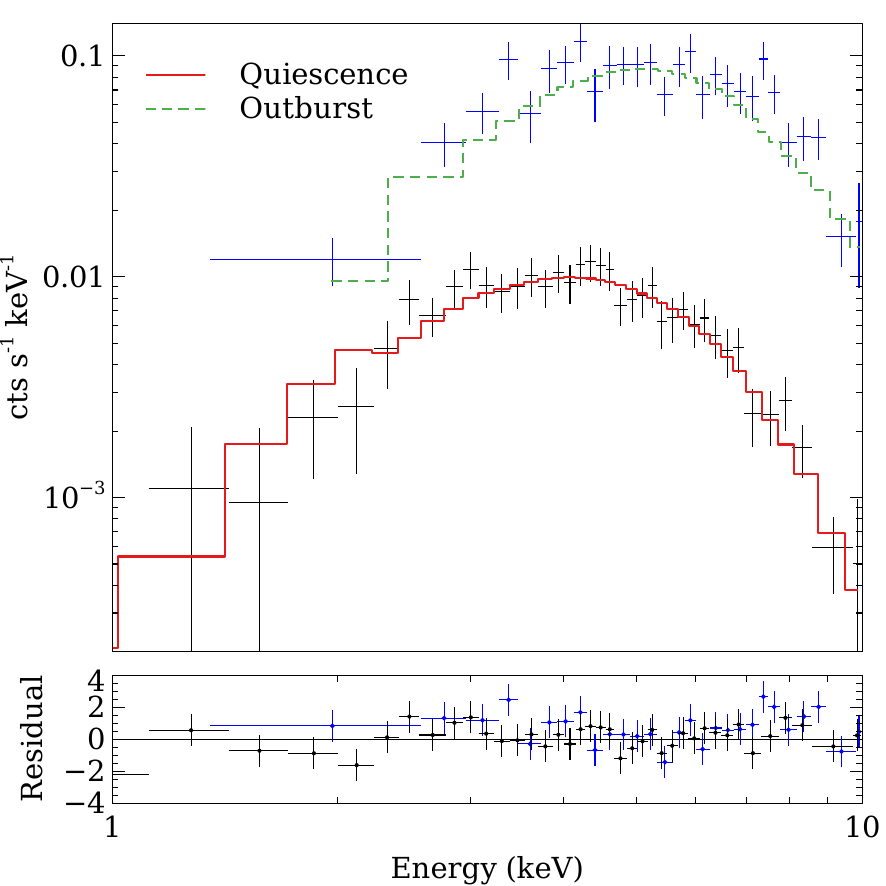}
\caption{Spectrum of the source 4XJ1751-2759 during the outburst period seen in observation 0886121001, and during the quiescent period throughout observation 0886080801. The black data points represent the spectrum of the source during observation 0886080801, and associated errors, as folded through the instrument response, and the solid red line represents the best fitting \texttt{tbabs$\times$bbodyrad} model. The blue data points represent the spectrum of the source during the outburst from observation 0886121001, and associated errors, as folded through the instrument response, and the dashed green line represents the best fitting \texttt{tbabs$\times$bbodyrad} model. For the purpose of this plot the spectrum has been rebinned to a minimum signal-to-noise of 5.0.}
\label{fig:spec_compare}
\end{figure}

The first result of note is that there is no apparent relationship between the blackbody temperature and the estimated flux of the source, other than that the temperature is greatest while the source was in outburst. We also note that there is no apparent relationship between the derived column density and the estimated flux. The temperature across all epochs is greater than $\sim1$\,keV, and the temperature for the outburst is greater than 10\,keV. We also note that the temperature while in outburst and during the faintest epoch, observation 0886120901, was poorly constrained. This is not unexpected during the outburst as \emph{Chandra} and \emph{XMM-Newton} are limited to only around 10 keV. For both observatories the effective areas begin to significantly decrease above $\sim$8\,keV. We present the spectra during the outburst and during one observation when it was in quiescence, in Figure \ref{fig:spec_compare}. Finally, if we fit to all five epochs simultaneously, requiring the column density to remain constant across all epochs we find the column density to be $\left(3.86^{+0.88}_{-0.75}\right)\times10^{22}\text{cm}^{-2}$, which is well above that determined by \cite{hi4pi_collaboration_hi4pi_2016} of $1.24\times10^{22} \text{cm}^2$. This simultaneous fit also has a consistent temperature within errors across the four epochs when not in outburst of $\sim2$\,keV.

We also consider the spectrum during the individual flares, as identified in Section \ref{subsec:outburst}, however, there are insufficient counts during all but the first flare to allow for the creation and fitting of robust spectra. During the first flare we find that the source reaches an unabsorbed flux of $3.94\times10^{-11} $erg cm$^{-2}$ s$^{-1}$ with a blackbody temperature of 100.0\,keV and column density of $1.46\times10^{22} \text{cm}^{-2}$. This temperature is, however, unconstrained, just as it is when all flares are considered together as one outburst in observation 0886121001, and the goodness of fit is $\chi^2_{\nu}=0.510$ for 6 degrees of freedom.

\subsection{Flux evolution}
\label{subsec:evol_flux}

\begin{figure}
\centering
\includegraphics[width=\columnwidth]{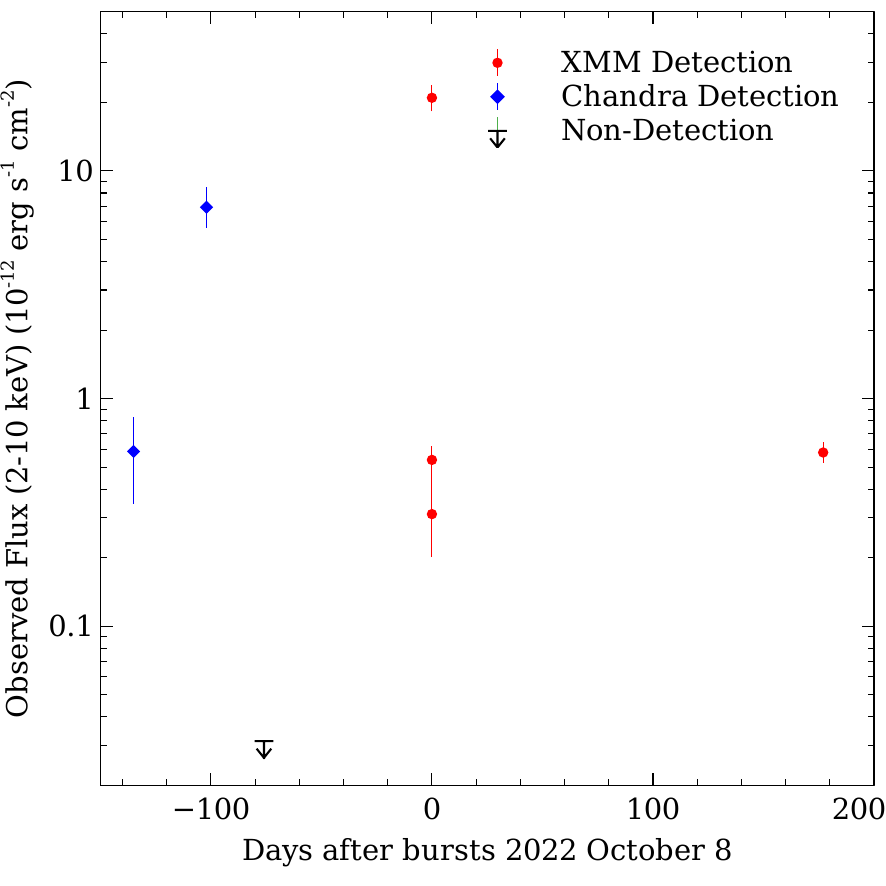}
\caption{Flux in the 2.0--10.0\,keV energy band over time, relative to the observed outburst on 8 October 2022. Fluxes are either reported as observed fluxes with 90\% confidence intervals, or as an upper limit on the \emph{Chandra} flux when not detected. Source detections with \emph{XMM-Newton} and \emph{Chandra} are displayed as red circles and blue diamonds respectively.}
\label{fig:flux_evol}
\end{figure}

We now consider the evolution of the X-ray flux for the source during the one-year period which covers the six observations of the source with \emph{Chandra} and \emph{XMM-Newton}. For the one observation where the source was not detected by \emph{Chandra} the 90\% upper limit on the flux was estimated using the \emph{Chandra} analysis task \texttt{aplimits}. We determine the observed flux of the source, with 90\% confidence limits, for the five observations where the source was detected. These values, with 90\% confidence intervals are shown in Figure \ref{fig:flux_evol}.

In Figure \ref{fig:flux_evol} we can see that the flux of the source varies by up to three orders of magnitude over the period of interest, approximately one year, and can change by up to two orders of magnitude within the space of a few hours. The fluxes during three of the six observations, and also during the quiescent phase at the start of observation 0886121001 preceding the outburst, are all within the range from $\sim 0.2-1 0\times 10^{-12} \text{\,erg s}^{-1} \text{cm}^{-2}$, but are not constant within the determined errors. We also note that observation 24193 has a significantly higher flux level than the four other observations which also do not contain outburst activity. This high flux is consistent with the flux measured when it was detected with the \emph{XMM-Newton} slew survey (see Fig. \ref{fig:flux_evol2}). The flux during this slew observation, from 30th March 2017, is estimated to be $9.4\times 10^{-12} \text{\,erg s}^{-1} \text{cm}^{-2}$. The remaining nine \emph{XMM} slew observations, where 4XJ1751-2759 was not detected, cover the period from September 2007 to June 2023, and the 2$\sigma$ upper limits on the estimated fluxes span nearly one order of magnitude from $\sim2.9-22\times10^{-12} \text{\,erg s}^{-1} \text{cm}^{-2}$. Given that these upper limits are greater than the fluxes at which we have detected the source in pointed observations, we cannot necessarily draw any robust conclusions from the non-detections during these times.

\subsection{Search For X-ray Pulsations}
\label{subsec:pulse_search}

We searched for pulsations in all \emph{XMM} and \emph{Chandra} observations where the source was detected. As described in Sec. \ref{sec:intro}, magnetars often show pulsations in the $\sim$2-10 keV energy range, and therefore detecting similar pulsations would provide further evidence for the magnetar interpretation. We used the approaches outlined in Section \ref{subsec:methods_data-temporal} to search for pulsations in un-binned and binned data in both the time and Fourier domains.

For the time resolutions of 100ms and 5s for \emph{XMM} and \emph{Chandra} data respectively, we do not find any significant features in the power spectra for any of the observations examined. We also do not find significant features in the averaged power spectra when averaging over good time interval segments of 50\,s, 100\,s, or 250\,s. There are no significant features identified when the quiescent and outburst sections of observation 0886121001 are taken separately. We also bin the event data down to the \emph{XMM} EPIC pn frame rate of 73.4\,ms and create power spectra but find no significant features in the lightcurves. Follow-up of minor features in the power spectra using either Epoch Folding or $Z^2_n$ searching does not find any significant pulsations when conducted using \texttt{stingray} or \texttt{HENDRICS} \citep{bachetti_hendrics_2018}. An accelerated search using \texttt{HENaccel} found no pulsations. Finally, we conducted a broad search for pulsation using the $Z^2_n$ searching algorithm in \texttt{HENDRICS} in the range from 0.01--25.0\,Hz, with values for n up to, and including, 15. This search was implemented for three epochs where the count rates were highest and we could minimise the effects of instrumental frame times. We examine the full observations 0886121001 and 0886080801, and also the period during the outburst in observation 0886121001 separately. The searches during these epochs gave 90\% upper limits on the pulsation fraction in each case as 36.1\%, 36.0\%, and 48.6\% at frequencies of 9.36\,Hz,  12.72\,Hz, and 1.77\,Hz respectively for the full observations 0886121001 and 0886080801, and the outburst during observation 0886121001. Further searches for pulsations using a narrower range of energies, 3.0--12.0\,keV, where the source signal is strongest against the background also produced no conclusive positive results.

\section{Discussion}
\label{sec:discuss}

The temporal variability and spectral features of 4XJ1751-2759 bear striking similarities to those of magnetar outbursts. The rapid rise in flux over a period of $\sim$100\,s, accompanied by a spectral hardening, followed by a decay over a period of months is in agreement with the temporal and spectral evolution seen in outbursts in confirmed magnetar sources \citep{rea_outburst_2013,coti_zelati_systematic_2018}. The absence of a clear optical or IR counterpart to the source further strengthens this classification. If we consider the case of any source not being detected by \emph{GAIA} at an apparent magnitude of 20.7 we find a limiting optical flux of $5.11\times10^{-14}\text{erg s}^{-1}\text{cm}^{-2}$, using the approach of \cite{schmitt_nature_2022}. As such 4XJ1751-2759 would have a lower limit on its optical to X-ray flux ratio of $log(F_\text{X} / F_\text{Opt})\sim$2.77 when at its brightest. This is consistent with that of known magnetars \citep[][etc.]{durant_search_2008,durant_search_2011}, and would not be consistent with a stellar interpretation. The absence of an observable optical or IR counterpart is not uncommon, and 15 of the 31 known magnetars either have no known counterpart or only have upper limits on the magnitude of a potential counterpart \citep{olausen_mcgill_2014}. The location of 4XJ1751-2759 is also consistent with those of known magnetars, in the direction of the Galactic Centre. For those with known distance estimates, at least 13 lie in the direction of the Galactic Centre \citep{kaspi_magnetars_2017}, and corresponding heights from the Galactic Plane range up to $\sim$180\,pc. The height below the plane, assuming 4XJ1751-2759 is located in the Galactic centre at a distance of 8\,kpc in line with similar distance estimates to known magnetars \citep{olausen_mcgill_2014}, of $\sim$90pc, is consistent with the heights of known magnetars. There is a supernova remnant just over one degree from the location of 4XJ1751-2759, G1.9+0.3, which has no associated central compact object (CCO). The age of the supernova is, however, only around 100 years, and as such the distance to 4XJ1751-2759 is too great for such an association. No pulsations were detected for 4XJ1751-2759, which may be because they are shorter than the frame time, too faint to be detected or non-existent.

\begin{figure}
\centering
\includegraphics[width=\columnwidth]{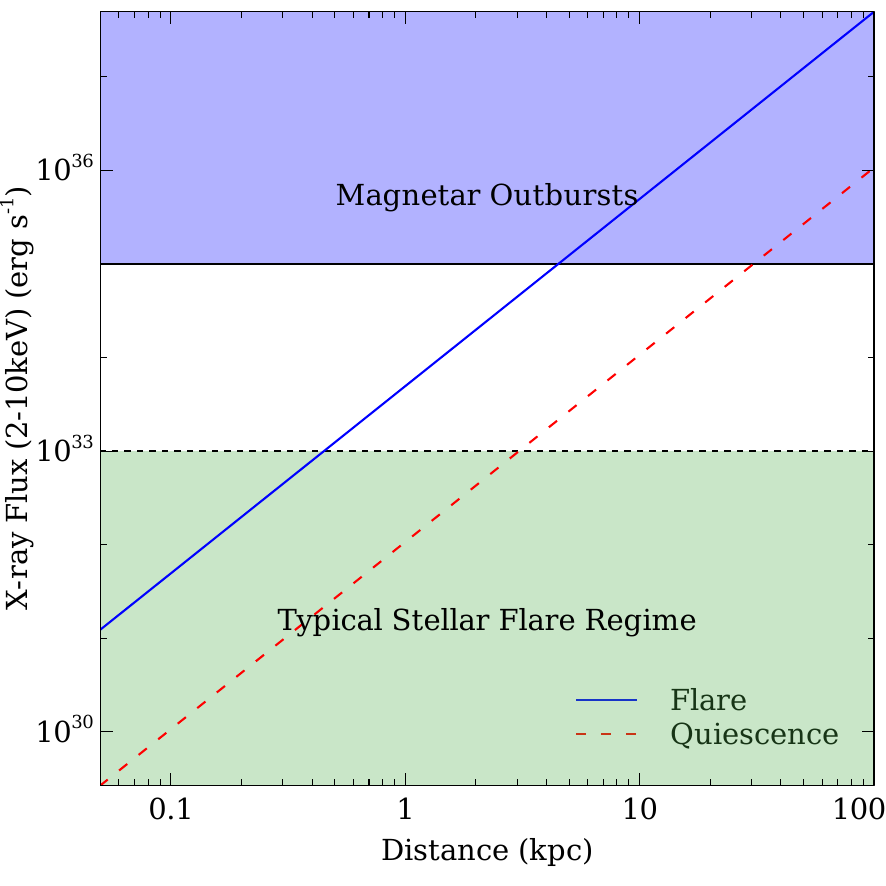}
\caption{Estimated flux of 4XJ1751-2759 in the 2.0-10.0\,keV band as a function of distance. We overlay the typical fluxes seen in stellar flaring activity and during magnetar outbursts. The solid blue line shows the flux during the outburst detected on 8th October 2022, and the dashed red line shows the flux during the quiescent period before the outburst.}
\label{fig:flux_lum}
\end{figure}

\begin{figure}
\centering
\includegraphics[width=\columnwidth]{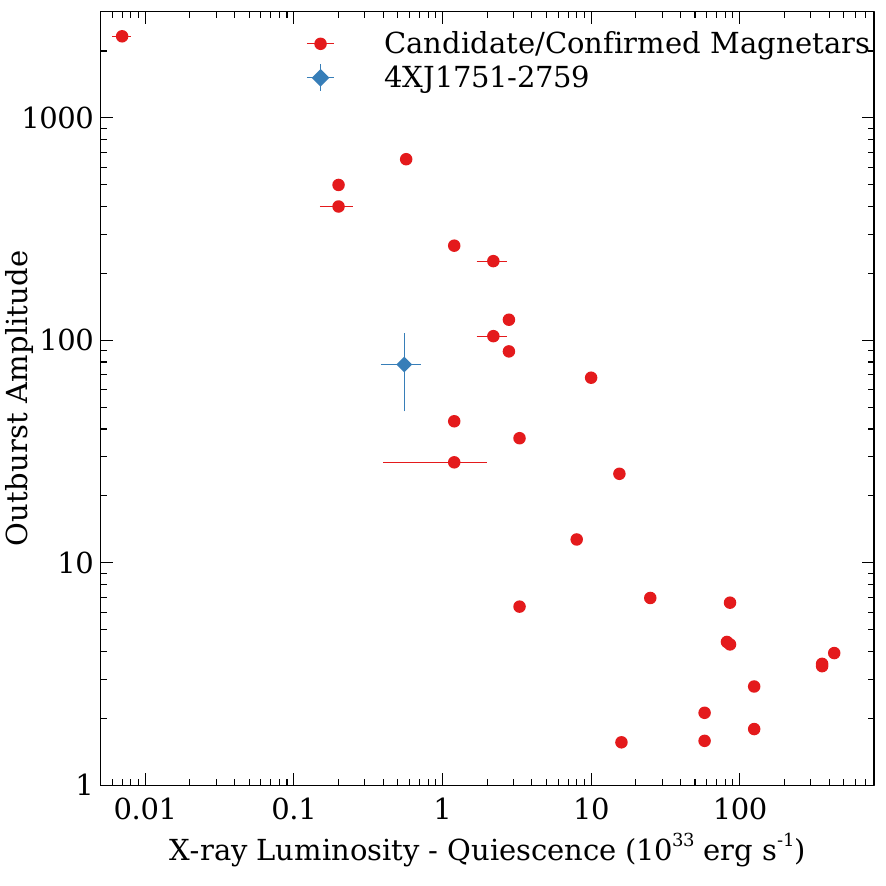}
\caption{Quiescent X-ray luminosity for confirmed and candidate magnetar sources, and 4XJ1751-2759 in comparison with the proportional increase in X-ray luminosity during outbursts. The outburst details for known or candidate magnetars are plotted as red circles, using the data in Tables 4 and 5 of \protect\cite{coti_zelati_systematic_2018}, and as an adaptation of Figure 3 of the same. The known outburst details for 4XJ1751-2759 are plotted as a blue diamond with errors, as the average (and standard deviation) of the quiescent luminosities taken from the \emph{XMM-Newton} observations 0886120901, 0886121001 and 0886080801 and in comparison with the peak luminosity during the outburst on the 8th October 2022. The luminosities for 4XJ1751-2759 are calculated assuming a distance of 8\,kpc.}
\label{fig:qu_amp_comp}
\end{figure}

The derived value for the peak flux during the brightening at the end of observation 0886121001 of $\sim3\times10^{-11} \text{\,erg s}^{-1} \text{cm}^{-2}$ implies a luminosity of $\sim2\times10^{35} \text{\,erg s}^{-1}$ at a distance of 8\,kpc. This luminosity is within the range of $(0.16-17)\times10^{35} \text{\,erg s}^{-1}$, see Fig. \ref{fig:flux_lum}, as seen in the outbursts in confirmed magnetars \citep{coti_zelati_systematic_2018}, and is still in agreement with the limiting luminosity for magnetar outbursts of $\sim10^{36}\text{\,erg s}^{-1}$ as observed by \cite{coti_zelati_systematic_2018}. Additionally, 4XJ1751-2759 also fits with the observed anti-correlation between quiescent X-ray luminosity and outburst amplitude seen in other magnetar sources. In Figure \ref{fig:qu_amp_comp} we show the comparison of quiescent X-ray luminosity against the proportional increase in luminosity during an outburst for known and candidate magnetar sources. That 4XJ1751-2759 has not yet been detected at higher energies, even during this outburst, does not make it an outlier in the population of magnetars. From the seventeen sources which have been observed in magnetar-like outburst, as listed in \cite{coti_zelati_systematic_2018}, four of the sources (XTE J1810-197, 1E 1048.1-5937, PSR J1119-6127, PSR J1846-0258) were detected as entering outburst through observations of their soft X-ray emission without additional detections at higher energies. The constraining non-detection with \emph{Chandra} on 24th July 2022 implies that the true quiescent emission from 4XJ1751-2759 is below that seen in all five of the other observations and that it is therefore in a brighter state during these epochs. Further, if the source has remained in an elevated state during the period from the outburst on 8th October 2022 through to 3rd April 2023, triggered by the observed outburst, then the timescales for variability would agree with the long-persistent tail observed in magnetar outbursts \citep{coti_zelati_systematic_2018} on scales of weeks-years. During the lower luminosity periods when the source is still detectable, an unabsorbed flux of $\sim0.7\times10^{-12}\text{\,erg s}^{-1}\text{cm}^{-2}$ at a distance of 8\,kpc, near the Galactic centre, implies an X-ray luminosity of the order of $\sim4 \times 10^{33}\text{\,erg s}^{-1}$. Such a luminosity is well within the ranges observed for the X-ray luminosity of other magnetar sources and candidates when not in outburst, which range from $\sim10^{30}-10^{35}\text{\,erg s}^{-1}$ \citep[Table 1 of ][]{kaspi_magnetars_2017}. The apparent non-detection of the source within one of the \emph{Chandra} observations could also imply that the source is transient in nature, like several other magnetars e.g. XTE 1810-197 \citep{gotthelf_imaging_2004}.

The spectral shape, being well described by an absorbed blackbody, or a combination of blackbody components, is in agreement with the spectra of magnetars and magnetar candidates. The temperature of the source, as determined through the spectral fitting in Section \ref{subsec:evol_spec} is, however, significantly higher than those seen in other magnetar sources when fit with similar models. The addition of secondary blackbody components, or of power law components, does not improve the quality of the fits in all cases, as shown in Table \ref{tab:spec_fits_all}. We also fit a non-thermal cut-off power law model during the outburst but find no improvement over a standard power law model during that epoch with an unconstrained cut-off energy (see Table \ref{tab:spec_fits_all}). For magnetar sources during quiescence, temperatures have been determined at $kT \lesssim 0.7$\,keV \citep{kaspi_magnetars_2017}. We find temperatures across all epochs of greater than 1\,keV, but the temperatures determined by a two blackbody fit, shown in Table \ref{tab:spec_fits_all}, to observations 24193 and 0886080801 are consistent within errors to the "warm" and "hot" blackbody components in a three-component spectrum of the magnetar XTE J1810-197 several months after an observed outburst \citep{borghese_x-ray_2021}. This could indicate the existence of an unobserved outburst some time before the \emph{Chandra} observation 24193. The derived temperatures of greater than 1\,keV do not therefore preclude a magnetar classification if 4XJ1751-2759 is not in quiescence during these epochs, consistent with a quiescent level below detection limits. The radii of the blackbody regions determined outside the outburst are small when compared to the size of neutron stars, in the range from 0.02--0.21km. Small radii are always determined when fitting a black body to a neutron star as the emission passes through a hydrogen atmosphere, so the black body fit is poorly adapted to derive the radius \citep{zavlin_model_1996}. The radii derived are, however, no smaller than some of those determined by hotspot models to other magnetars in the time around outbursts \citep[e.g.][]{rea_outburst_2013}, and may be greater if the source is more distant than the galactic centre. To try and further constrain higher energy emission, and the potential spectral shape, from 4XJ1751-2759 we searched for any detection with the \emph{Fermi} Gamma-Ray Burst Monitor. There are no recorded triggers (above or below threshold) at the time and location of the outburst, and a targeted search \citep{goldstein_updates_2019} in a window of 25\,s around the start of the outburst finds no sub-threshold trigger candidates which could be coincident with the position of 4XJ1751-2759. As such we instead define an upper limit on the hard X-ray flux in the range from 8.0--30.0\,keV of $(0.9-11.7)\times10^{-8}$erg\,cm$^{-2}$\,s$^{-1}$, depending on the spectral template used, at the location of 4XJ1751-2759. By extrapolating the \emph{XMM-Newton} spectrum during outburst the out to 30\,keV we would expect a flux in this range of the order 10$^{-10}$\,erg\,s$^{-1}$\,cm$^{-2}$, and so the \emph{Fermi} upper limit does not place any constraints on the spectral shape at higher energies. We would not expect \emph{Fermi} to be able to detect this source.

The closest possible counterpart to the source, and the only source in optical or infra-red catalogues within 5$\sigma$ of the X-ray source position and error, is the GAIA source 4057461686227145088, which is coincident on the sky with the 2MASS source 17513680-2758583. It has apparent magnitudes $m_G = 19.60$ and $m_J = 13.42$ in the optical and near-IR bands respectively. It is located at a separation of \ang{;;1.54} (3.21$\sigma$) from the X-ray source position and has been determined to be at a distance of between 5.5 and 8.7\,kpc as the 16-84\% interval on the photogeometric distance to the source \citep{bailer-jones_estimating_2021}, or further still based on the parallax of 0.0597mas at a distance of 16.8kpc. The error on this parallax is very large (0.9191), in comparison to the measured parallax, however, and so we instead use the smaller distances from the photogeometric analysis which give more conservative estimates on the X-ray luminosity. This object has no associated temperature estimate or stellar type classification in Gaia catalogues. We therefore fit the source using ARIADNE \citep{vines_ariadne_2022} and find that the best fitting spectral type is of a G-type star with $T_{\text{eff}} \sim5900\,$K and with a lower bound on its distance estimate $d_{\text{pc}} \sim 7.7\,\text{kpc}$. Considering these distances the X-ray luminosity of the source during its outburst would be in the range from $0.99-2.44 \times 10^{35}$\,erg\,s$^{-1}$, higher, by a factor of $\sim$100, than the accepted ranges for stellar flaring activity \citep[][etc.]{maehara_superflares_2012,tsuboi_large_2016,kowalski_stellar_2024}, and with a quiescent X-ray luminosity in the range $\sim2.6-6.4 \times 10^{33}$, which is within the ranges predicted for transient magnetars \citep{kaspi_magnetars_2017}, as can be seen in Fig. \ref{fig:flux_lum}. Furthermore, GAIA 4057461686227145088 has a g-band magnitude of 19.60, and with a zero-point magnitude of 25.69 this gives an optical flux of $1.13\times10^{-13}\text{\,erg s}^{-1}\text{cm}^{-2}$.  This ratio also makes a stellar classification even less likely as the ratio for stellar objects is such that $log(F_\text{X} / F_\text{Opt})\leq$-0.4 \citep{maccacaro_x-ray_1988,huang_study_2010}, and instead strengthens the argument that it is a compact object. The spectral fitting also does not favour a stellar interpretation of 4XJ1751-2759. Single or double \texttt{apec} models offer fits to the observed spectra which are not superior to those of the blackbody model (see Table \ref{tab:spec_fits_all}), and the derived values for the temperatures and column densities are not physically consistent in several cases.

If we instead consider the possibility that the source is too faint to be optically detected by Gaia, as could be the case for a faint dwarf star we find that the flux is inconsistent with an undetected faint stellar source. Taking a conservative limit for the Gaia detection limit of magnitude 20.7 we would require even a star with absolute magnitude of 16 to be at a distance of at least 100pc. At this distance the peak X-ray luminosity would still be of the order $\sim 10^{32} \text{erg s}^{-1}$, as per Fig. \ref{fig:flux_lum}, and as such would be greater than the luminosities of such flares from the faintest dwarf stars by factors of at least $\sim10 - 100$ \citep{joseph_simultaneous_2024}. For more luminous stars, like those considered by \cite{pye_survey_2015}, to be undetected would require distances of the order of kiloparsecs, resulting in an X-ray luminosity for 4XJ1751-2759 $L_X \geq 10^{34} \text{erg s}^{-1}$. Again, such a luminosity is inconsistent with observed flares from stellar objects with these stellar types, see Fig. \ref{fig:flux_lum}, and is at least one order of magnitude greater than previously detected flares \citep[][etc.]{maehara_superflares_2012,pye_survey_2015,tsuboi_large_2016,kowalski_stellar_2024}.  This, in combination with the poor fits to spectra for plasma (\texttt{apec}) and bremsstrahlung models (see Tab. \ref{tab:spec_fits_all}) allows us to exclude stellar objects as the source of this X-ray emission. In addition, the column density obtained from fitting the combined X-ray spectrum is about three times the column density estimated by the \cite{hi4pi_collaboration_hi4pi_2016}, which may imply that the source is situated behind the Galactic bulge, which would be consistent with the highest distant estimate from the photogeometric estimate, implying a luminosity of $2\times10^{35}$ ergs s$^{-1}$, making it a factor 100 greater than the greatest expected luminosity of a star. This would further increase the luminosity of the source,  and would put the X-ray luminosity of the source further out of the possible range of stellar activity and even more firmly within magnetar ranges.

The temporal, energetic, and spectral features of 4XJ1751-2759 are similar to those seen in the phenomena of Supergiant Fast X-ray Transients (SFXTs) observed in high mass X-ray binaries \citep[e.g.][]{sidoli_supergiant_2017,fornasini_high-mass_2023}. By considering the source flux, an SFXT \citep[for a review see e.g.][]{sidoli_supergiant_2017} interpretation can also be ruled out, as the supergiant companion star would be detected out to 20 kpc. This is considerably larger than the maximum photogeometric distance estimate for the potential \emph{Gaia}/\emph{2MASS} counterpart. The average X-ray luminosity of SFXTs is of the order $\leq 10^{34}$\, erg s$^{-1}$ \citep{sidoli_supergiant_2017}. With a quiescent flux of the order of $10^{-12} \text{\,erg s}^{-1} \text{cm}^{-2}$ this would place 4XJ1751-2759 at a distance of no greater than 9.2\,kpc, see Fig. \ref{fig:flux_lum}, even for the brightest SFXTs. That the source also becomes significantly harder during its outburst allows us to exclude a possible classification as a low mass X-ray binary \citep[e.g.][]{remillard_x-ray_2006,munoz-darias_black_2014}. 

Finally, we consider the possibility that this source was a rotationally powered pulsar (RPP), which would explain the large $F_X/F_{Opt}$, and could explain why pulsations are not detected if they were of a shorter period than the instrumental frame rates. The spin-down luminosity of the source is determined as $L = 4\pi^2I\dot{P}/P^3$, and $I\approx10^{45}\text{g cm}^{-2}$ is the stellar moment of inertia, our observed luminosities would require $\dot{P}/P^3\sim1\times10^{-13}$. For typical pulsar periods and period derivatives this places the observed quiescent X-ray luminosity of the same order as, or greater than that of, the spin-down luminosity. For typical rotationally powered pulsars the X-ray luminosity is a small proportion of the spin-down luminosity. Further, should the source be at a greater distance than the Galactic centre, as is possibly implied by the higher estimated column density than the Galactic value, then the X-ray luminosity would increase further still. The thermal emission from RPPs also typically shows a softer spectrum and temperatures well below those seen in 4XJ1751-2759, $\sim0.1-0.3$\,keV \citep[e.g.][]{becker_x-ray_1997,rigoselli_new_2018,webb_thermal_2019,rigoselli_thermal_2022}. This makes classification of 4XJ1751-2759 as a rotationally-powered pulsar even less feasible. 

Assuming that this object is a Galactic X-ray source, the most likely candidate class for this combination of flux and luminosity ranges, temporal features and spectral temperatures is a magnetar, even without detectable pulsations.

\section{Conclusions}
\label{sec:concs}

We have presented a spectral and temporal analysis of six observations of 4XJ1751-2759 obtained with \emph{XMM-Newton} and \emph{Chandra}. We find that the observational characteristics are more resemblant of a magnetar than of other possible X-ray source classifications, even in the absence of significant pulsations in the X-ray event data.

   \begin{enumerate}
        \item 4XJ1751-2759 can be well described, when not in outburst, by an absorbed blackbody with a temperature of $\sim2-4$\,keV. The temperature during outburst is estimated at $\sim10$\,keV, but not able to be constrained with available observations.
        \item The estimated column density to 4XJ1751-2759 is greater than the Galactic value at the distance of the only close optical/IR counterpart, possibly placing the source at a significant distance, greater than the Galactic centre implying peak X-ray luminosities $L_{X} \geq2\times10^{35}\text{\,erg s}^{-1}$.
        \item There are no detectable pulsations in the X-ray data, but this may be due to the low count rate.
        \item The estimated luminosity of 4XJ1751-2759 during quiescent phases is of the order $\sim4\times10^{33}\text{\,erg s}^{-1}$, greater than that of ordinary stellar objects. 
        \item The X-ray-to-optical flux ratio, $log(F_\text{X} / F_\text{Opt})\sim$2--3, is well beyond stellar ranges ($log(F_\text{X} / F_\text{Opt})\leq$-0.4) and consistent with that in magnetar outbursts.
        \item There is no clear supergiant optical or IR companion to 4XJ1751-2759 making classification as a supergiant fast X-ray transient impractical.
   \end{enumerate}

Given this evidence a magnetar candidate classification appears a more likely explanation for the variability of 4XJ1751-2759 than that of other known X-ray transient classes, and any potential future detection of pulsations would definitively confirm this. In order to make such a conclusive classification of 4XJ1751-2759 many more counts are required to detect pulsations and subsequently perform robust $P$ and $\dot{P}$ estimation. Such information would also allow us to estimate the strength of the magnetic fields around 4XJ1751-2759, and make a determination of the lifetime of 4XJ1751-2759. In the absence of further data we cannot rule out the possibility either, that 4XJ1751-2759 could be the first example of a new class of X-ray transient. The addition of concurrent higher energy observations during any subsequent outbursts would also allow for constraints to be placed on the temperature during these very bright phases.

\section*{Acknowledgements}

Based on observations obtained with XMM-Newton, an ESA science mission with instruments and contributions directly funded by ESA Member States and NASA. This research has made use of ESASky, developed by the ESAC Science Data Centre (ESDC) team andiscd maintained alongside other ESA science mission's archives at ESA's European Space Astronomy Centre (ESAC, Madrid, Spain). R. Webbe and N. Webb are grateful to the Centre National d’Études Spatiales (CNES) for their outstanding support for the XMM-SSC activities. This project has received funding from the European Union’s Horizon 2020 research and innovation programme under grant agreement Number 101004168.
All figures in this work were created with \texttt{Veusz} and \texttt{matplotlib}. The authors thank the anonymous reviewers for their constructive feedback.

%%%%%%%%%%%%%%%%%%%%%%%%%%%%%%%%%%%%%%%%%%%%%%%%%%
\section*{Data Availability}

All observational data used in this analysis are freely available through the \emph{XMM-Newton} Science Archive (\url{https://www.cosmos.esa.int/web/xmm-newton/xsa}) and the \emph{Chandra} Data Archive (\url{https://cxc.harvard.edu/cda/}). 

The details of all analysis leading to the results reported herein are part of the Git repository \url{https://github.com/robbie-webbe/4XJ1751_magnetar} .



\bibliographystyle{mnras}
\bibliography{magnetar_cand_4XJ1751} % if your bibtex file is called example.bib

%%%%%%%%%%%%%%%%%%%%%%%%%%%%%%%%%%%%%%%%%%%%%%%%%%

%%%%%%%%%%%%%%%%% APPENDICES %%%%%%%%%%%%%%%%%%%%%

\appendix
\onecolumn

\section{Spectral Fitting}
\label{app:spec_fits}
We list here, in full, the results of all power spectral models fitted to the five epochs considered during this analysis. This supplementary table is designed to sit alongside and complement Table \ref{tab:spec_fits} where we show only the results of the fitting with a thermal blackbody model in comparison with the spectra of confirmed magnetars, and the results and discussion put forth in Sections \ref{subsec:evol_spec} and \ref{sec:discuss} as to the possible classifications of the source 4XJ1751-2759. For those models which utilise multiples of the same model, \texttt{tbabs*(bbody + bbody)} and \texttt{tbabs*(apec + apec)} we initialise the two models with component temperatures of 0.5\,keV and 4.0\,keV respectively. We report the 90\% uncertainties on parameters, and where the fit is insufficiently sensitive to a parameter to allow this we report the uncertainty in that direction as $+ \infty$ or $- \infty$ to indicate that it is not appropriately constrained by the fitting.

\begin{table*}
\renewcommand{\arraystretch}{1.5}
\caption{Spectral fits to observations of 4XJ1751-2759, and associated 90\% confidence intervals on spectral parameters. The last two columns show the number of bins used in the fit $N$, and the Cash statistic for the optimal fit $C$.}
\label{tab:spec_fits_all}
\centering
\begin{tabular}{cccccccc}
\hline\hline
Epoch & Model & nH & $kT$ & $\alpha$ & $kT_2$ & $N$ & $C$ \\
 & \texttt{tbabs $\times$} & ($10^{22}$ cm$^{-2}$) & (keV) &  & (keV) & & \\
\hline
    \multirow{7}{6em}{24193} & \texttt{bbody} & $5.72^{+2.14}_{-1.73}$ & $1.80^{+0.53}_{-0.36}$ & -- & -- & \multirow{7}{*}{276} & 215.2 \\
    & \texttt{powerlaw} & $9.14^{+3.09}_{-2.58}$ & -- & $1.46^{+0.71}_{-0.63}$ & -- &  & 210.6 \\
    & \texttt{bremss} & $8.65^{+2.40}_{-1.58}$ & $54.54^{+\infty}_{-\infty}$ & -- & -- &  & 214.4 \\
    & \texttt{apec} & $8.68^{+1.51}_{-1.36}$ & $40.72^{+\infty}_{-27.03}$ & -- & -- &  & 211.3 \\
    & \texttt{(bbody + bbody)} & $16.76^{+7.99}_{-6.96}$ & $0.40^{+0.29}_{-0.12}$ & -- & $2.30^{+4.43}_{-0.79}$ &  & 204.3 \\
    & \texttt{(bbody + po)} & $22.17^{+7.47}_{-9.11}$ & $2.45^{+\infty}_{-0.95}$ & $7.58^{+\infty}_{-4.01}$ & -- &  & 204.0 \\
    & \texttt{(apec + apec)} & $25.15^{+7.02}_{-6.90}$ & $0.41^{+47.699}_{-0.09}$ & -- & $64.00^{+\infty}_{-43.53}$ &  & 209.3 \\
\hline
    \multirow{7}{6em}{0886120901} & \texttt{bbody} & $0.92^{+27.08}_{-\infty}$ & $3.98^{+\infty}_{-\infty}$ & -- & -- & \multirow{7}{*}{211} & 187.4 \\
    & \texttt{powerlaw} & $1.77^{+25.94}_{-\infty}$ & -- & $-0.15^{+8.62}_{-1.12}$ & -- &  & 187.6\\
    & \texttt{bremss} & $9.68^{+31.39}_{-5.52}$ & $99.65^{+\infty}_{-\infty}$ & -- & -- &  & 188.2 \\
    & \texttt{apec} & $9.97^{+7.92}_{-5.15}$ & $40.43^{+\infty}_{-34.12}$ & -- & -- &  & 188.1 \\
    & \texttt{(bbody + bbody)} & $0.88^{+27.11}_{-\infty}$ & $0.00^{+\infty}_{-\infty}$ & -- & $4.03^{+\infty}_{-\infty}$ &  & 187.4 \\
    & \texttt{(bbody + po)} & $22.50^{+39.58}_{-\infty}$ & $200.0^{+\infty}_{-\infty}$ & $4.71^{+0.47}_{-\infty}$ & -- &  & 187.9 \\
    & \texttt{(apec + apec)} & $11.02^{+3.47}_{-2.67}$ & $0.11^{+0.09}_{-\infty}$ & -- & $38.52^{+\infty}_{-31.72}$ &  & 185.6 \\
\hline
    \multirow{7}{6em}{0886121001 (Quiescence)} & \texttt{bbody} & $0.59^{+1.29}_{-\infty}$ & $4.60^{+8.51}_{-1.82}$ & -- & -- & \multirow{7}{*}{435} & 333.6 \\
    & \texttt{powerlaw} & $1.02^{+1.85}_{-0.92}$ & -- & $-0.31^{+0.59}_{-0.49}$ & -- &  & 333.9 \\
    & \texttt{bremss} & $6.44^{+4.02}_{-2.33}$ & $200.00^{+\infty}_{-\infty}$ & -- & -- &  & 346.7 \\
    & \texttt{apec} & $7.66^{+5.23}_{-2.98}$ & $21.65^{+\infty}_{-10.1}$ & -- & -- &  & 345.9 \\
    & \texttt{(bbody + bbody)} & $0.59^{+1.29}_{-\infty}$ & $0.00^{+\infty}_{-\infty}$ & -- & $4.61^{+8.51}_{-1.82}$ &  & 333.6 \\
    & \texttt{(bbody + po)} & $1.02^{+3.07}_{-0.92}$ & $200.00^{+\infty}_{-\infty}$ & $-0.31^{+0.60}_{-0.49}$ & -- &  & 333.9 \\
    & \texttt{(apec + apec)} & $7.56^{+5.31}_{-2.92}$ & $21.64^{+\infty}_{-\infty}$ & -- & $64.00^{+\infty}_{-\infty}$ &  & 345.9 \\
\hline
    \multirow{8}{6em}{0886121001 (Outburst)} & \texttt{bbody} & $5.08^{+2.00}_{-1.62}$ & $10.50^{+\infty}_{-\infty}$ & -- & -- & \multirow{8}{*}{529} & 383.6 \\
    & \texttt{powerlaw} & $5.71^{+2.61}_{-2.09}$ & -- & $-0.62^{+0.46}_{-0.43}$ & -- &  & 383.3 \\
    & \texttt{bremss} & $15.01^{+1.84}_{-2.32}$ & $200.00^{+\infty}_{-\infty}$ & -- & -- &  & 416.7 \\
    & \texttt{apec} & $15.37^{+2.91}_{-2.35}$ & $64.00^{+\infty}_{-32.35}$ & -- & -- &  & 419.1 \\
    & \texttt{(bbody + bbody)} & $10.99^{+7.91}_{-4.26}$ & $0.67^{+9.58}_{-0.30}$ & -- & $200.00^{+\infty}_{-\infty}$ &  & 381.1 \\
    & \texttt{(bbody + po)} & $13.44^{+2.89}_{-9.42}$ & $200.00^{+\infty}_{-\infty}$ & $3.67^{+0.73}_{-\infty}$ & -- &  & 382.3 \\
    & \texttt{cutoffpl} & $5.69^{+2.61}_{-3.07}$ & $500.00^{+\infty}_{-499.99}$ & $-0.63^{+0.46}_{-2.07}$ & -- &  & 383.3 \\
    & \texttt{(apec + apec)} & $33.00^{+11.96}_{-12.02}$ & $0.33^{+0.12}_{-0.09}$ & -- & $64.00^{+\infty}_{-28.86}$ &  & 409.6 \\
\hline
    \multirow{7}{6em}{0886080801} & \texttt{bbody} & $3.03^{+1.12}_{-0.91}$ & $1.91^{+0.30}_{-0.24}$ & -- & -- & \multirow{7}{*}{750} & 550.8 \\
    & \texttt{powerlaw} & $6.05^{+1.75}_{-1.45}$ & -- & $1.33^{+0.40}_{-0.37}$ & -- &  & 549.1\\
    & \texttt{bremss} & $5.90^{+1.42}_{-0.94}$ & $78.37^{+\infty}_{-62.45}$ & -- & -- &  & 549.1 \\
    & \texttt{apec} & $5.92^{+1.06}_{-0.74}$ & $64.00^{+\infty}_{-43.94}$ & -- & -- &  & 549.9 \\
    & \texttt{(bbody + bbody)} & $9.37^{+5.35}_{-5.35}$ & $0.42^{+0.18}_{-0.07}$ & -- & $1.95^{+0.89}_{-0.38}$ &  & 546.3 \\
    & \texttt{(bbody + po)} & $12.88^{+6.36}_{-7.79}$ & $2.03^{+\infty}_{-0.46}$ & $6.03^{+3.10}_{-5.78}$ & -- &  & 546.4 \\
    & \texttt{(apec + apec)} & $6.39^{+2.94}_{-0.98}$ & $3.43^{+\infty}_{-\infty}$ & -- & $64.00^{+\infty}_{-39.63}$ &  & 547.6 \\
\hline
\end{tabular}
\end{table*}

\section{Long-Term Source Flux}
\begin{figure*}
\centering
\includegraphics[width=\textwidth]{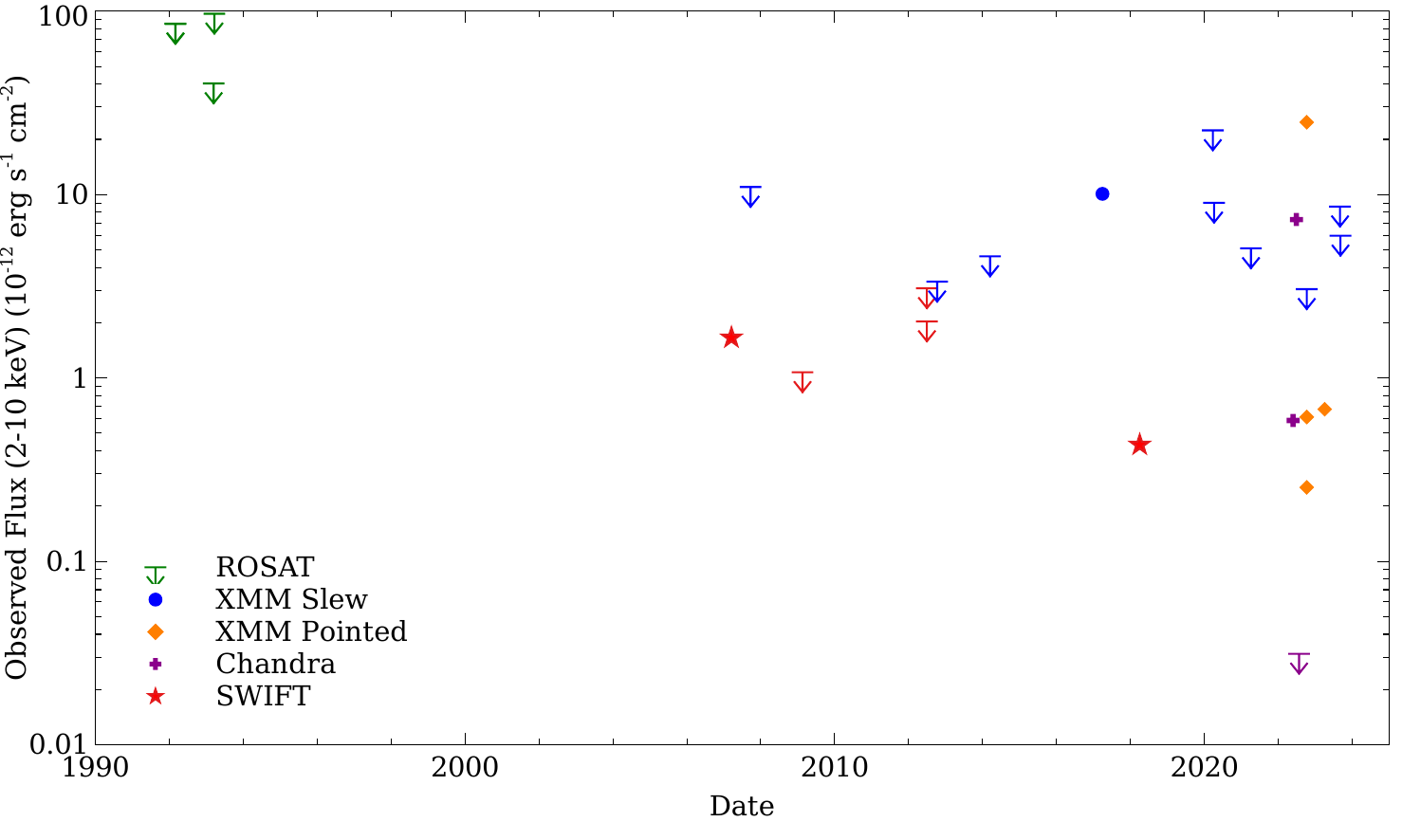}
\caption{Long term lightcurve, including upper limits on the flux from the source position for five X-ray observatories. Data is collected from \emph{ROSAT}, \emph{Swift} XRT, \emph{Chandra} and \emph{XMM} pointed and slew observations. Fluxes are plotted as blue circles, orange diamonds, purple crosses, and red stars as fluxes from \emph{XMM} slew, \emph{XMM} pointed, \emph{Chandra}, and \emph{Swift} XRT detected values respectively, and with arrows indicating upper limits on fluxes across all five observation types. All source fluxes are in the band from 2--10\,keV, and have been estimated from the provided count rate with a neutral hydrogen column density of $2\times10^{22}\text{cm}^{-2}$ and with a blackbody temperature of 2\,keV.}
\label{fig:flux_evol2}    
\end{figure*}

The position of 4XJ1751-2759 has been observed multiple times over a period of decades. In Figure \ref{fig:flux_evol2} we present the long term evolution of the source flux as observed using \emph{ROSAT}, \emph{Swift} XRT, \emph{Chandra} and \emph{XMM-Newton}. These flux values are taken from the \emph{XMM} HILIGT upper limit server\footnote{\url{http://xmmuls.esac.esa.int/hiligt/}} using their default 90\% significance level, and are converted from mission count rates in the bands provided to fluxes in the range from 2--10\,keV. For this flux conversion we assume a neutral hydrogen column density of $2\times10^{22}\text{cm}^{-2}$ and with a blackbody temperature of 2\,keV, as indicated by the spectral fitting in Section \ref{subsec:evol_spec}.

\section{Outburst Flare Structure}
\begin{figure*}
\centering
\includegraphics[width=\textwidth]{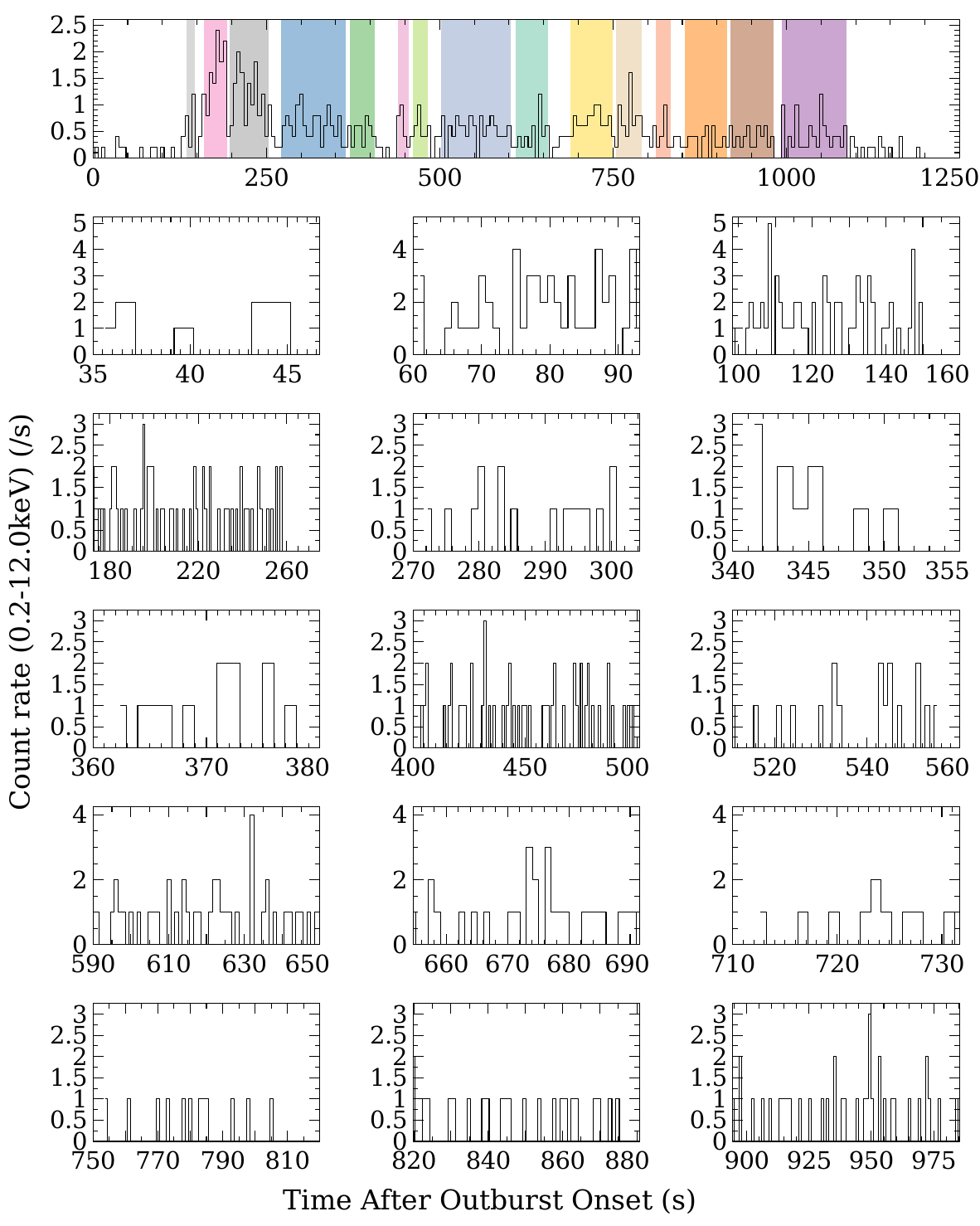}
\caption{Locations and structures of flares identified by HDBSCAN during the outburst during observation 0886121001. In the top panel the lightcurve bins are 5\,s, and for the lower panels the bins are 1\,s. All lightcurves are for X-ray energies in the range from 0.2--12.0\,keV.}
\label{fig:outburst_flares}    
\end{figure*}

We performed clustering using \texttt{HDBSCAN} on the photon event times during the outburst as described in Section \ref{subsec:methods_data-temporal} during observation 0886121001. For this purpose we cluster the events after the XMM time 781623400, corresponding to approximately 8 October 2022 13:35:30 UTC. This approach identified 15 clusters of photons arrival times. In Figure \ref{fig:outburst_flares} we show the outburst as detected by \emph{XMM-Newton} and the lightcurves during the 15 clusters as defined by the clustering. In all cases the lightcurves are for X-ray events in the energy range from 0.2--12.0\,keV.

%%%%%%%%%%%%%%%%%%%%%%%%%%%%%%%%%%%%%%%%%%%%%%%%%%

% Don't change these lines
\bsp	% typesetting comment
\label{lastpage}
\end{document}